\documentclass[11pt]{article}

\usepackage[margin=1in]{geometry}
\usepackage[T1]{fontenc}

\usepackage[utf8]{inputenc}
\usepackage[activate={true,nocompatibility},final,tracking=false,kerning=false,spacing=false,factor=1100,stretch=0,shrink=0]{microtype}
\usepackage{amsmath}
\usepackage{booktabs}
\usepackage{graphicx}
\usepackage{caption}
\usepackage[numbers,sort&compress]{natbib}
\usepackage[colorlinks=true,linkcolor=blue,citecolor=blue,urlcolor=blue]{hyperref}

\newcommand{\Cone}{C1}
\newcommand{\Ctwo}{C2}
\newcommand{\Cthree}{C3}

\title{Mitigating Fabrication in Multi-Stage LLM Pipelines for Hiring:\\
An Empirical Evaluation of Prompt Guardrails and\\ Human-in-the-Loop Checkpoints}

\author{Hiroko Takano\\
Lemmanode LLC\\
\texttt{contact@lemmanode.com}}

\date{July 2026}

\begin{document}
\maketitle

\begin{abstract}
Multi-stage LLM pipelines for hiring tasks such as resume improvement, interview question generation, and answer feedback can fabricate credentials, inflate qualifiers, and invent experience. We empirically evaluate two mitigations, prompt guardrails and human-in-the-loop (HITL) checkpoints between pipeline stages, against a fully automated baseline. In a controlled experiment (10 synthetic resumes $\times$ 2 job descriptions $\times$ 3 repetitions $\times$ 3 conditions; 180 runs; single generation model, fixed seed and parameters), a fully automated baseline (\Cone) produced at least one unsupported claim in 96.7\% of outputs (mean 6.80 findings per output). Prompt guardrails (\Ctwo) reduced finding density by 86\% (0.92/output) but 50.0\% of outputs still contained at least one fabrication, indicating that prompt-level mitigation alone is insufficient. A human checkpoint after the resume-improvement stage (\Cthree) eliminated all identity fabrications, reduced finding density by 59\% (6.88 to 2.82, $p = .022$, exact sign test on pair-level rates), reduced item-level fabrication from 96.7\% to 75.0\%, and cut capture of JD-embedded trap requirements from 47\% to 2\%, compared with 5\% under the guardrail. An exploratory analysis of multi-specialty resumes shows contamination-type alterations rising monotonically with the domain distance between the candidate's specialties; career changers and multi-specialty professionals are therefore the most exposed population. The reviewer's residual-error profile was systematic: flagrant fabrications (invented identities) were always caught, while subtle qualifier drops and plausible new claims survived review roughly half the time (54.5\% removal). Neither mitigation degraded the deliverable: claim retention exceeded 99\% under both interventions, and usefulness differences were small. The two interventions are effective in complementary ways: the guardrail eliminates unprompted absence-list hits and qualifier inflation at negligible cost, while the checkpoint provides near-categorical guarantees on the most severe failure types, namely invented identities and JD-baited claims. These results support a layered architecture combining guardrails with a human checkpoint, identify concrete requirements for reviewer-support tooling, and motivate the combined condition as a natural next experiment. A supplementary run with a newer-generation model (90.0\% baseline fabrication rate) suggests the problem is not resolved by model progress alone.
\end{abstract}


\section{Introduction}

LLM-based tools are increasingly deployed across hiring workflows: rewriting resumes against a target job description (JD), generating tailored interview questions, and giving feedback on candidate answers. These tools serve candidates directly and are equally operated by recruiters, who routinely advise candidates on resumes and interview preparation. The pipelines are multi-stage: the output of one LLM call becomes the input of the next. This architecture creates a specific risk profile: a fabrication introduced at an early stage, such as an invented certification in an improved resume, propagates through the pipeline, becoming the factual premise of interview questions that assume the fabricated experience, of feedback that reinforces it, and ultimately of a real interview.

The risk is not hypothetical. The benchmark's resumes deliberately carry no names or contact details as a privacy measure; the fully automated baseline responded by inventing them. Complete fictional name headers appeared unprompted in the pilot run, prompting a protocol update before the main experiment, and recurred in 6 of 60 baseline outputs of the main run. No fabricated identity details are reproduced in this paper.

A second motivation comes from practice. Professionals whose careers span multiple specialties, for example design and data analysis or law and localization, commonly observe that AI rewriting tools blend their domains, restating experience in one field as credentials in the adjacent one. The first author's own multi-specialty background repeatedly surfaced this failure mode. The benchmark therefore embeds a controlled specialty-distance gradient (\S\ref{sec:benchmark}), and we report an exploratory contamination analysis along it (\S\ref{sec:contamination}).

The intuitive mitigation, and the one most cheaply available to developers, is a prompt guardrail: instructing the model not to add unsupported facts. The alternative is structural: inserting a human checkpoint between stages, where a reviewer with access to ground truth verifies and corrects the output before it is consumed downstream. HITL is widely recommended in AI governance frameworks, but its effect size in multi-stage generation pipelines is rarely quantified, and it carries a real cost in reviewer time. This paper quantifies that effect.

\paragraph{Research questions.}
\begin{itemize}
  \item \textbf{RQ1}: At what rate does a typical (non-strawman) automated resume-improvement pipeline fabricate, and what kinds of fabrication occur?
  \item \textbf{RQ2}: How much fabrication does a prompt guardrail remove, and what residual remains?
  \item \textbf{RQ3}: How much fabrication does a human checkpoint remove, what residual remains, and what is the reviewer's error profile?
  \item \textbf{RQ4} (supplementary): Does a newer-generation model change the baseline picture?
\end{itemize}

\paragraph{Contributions.}
\begin{enumerate}
  \item A controlled 180-run comparison of automated, guardrail, and HITL conditions on a purpose-built, PII-free benchmark with claim-level ground truth and explicit absence lists (fabrication ``traps'').
  \item Quantified evidence that prompt guardrails reduce fabrication density by 86\% yet leave fabrication in half of outputs, and that a single human checkpoint removes all identity-level fabrication and 59\% of finding density.
  \item A characterization of what human review misses (qualifier drops, plausible additions), yielding design requirements for reviewer-support tooling.
  \item Documentation of reviewer protocol deviations and their handling, treating reviewer fallibility as a measured property of the checkpoint rather than as noise.
\end{enumerate}

\section{Related Work}

\paragraph{Hallucination and fabrication in LLM generation.}
Hallucination, that is, fluent but unsupported or false generation, is extensively documented; Huang et al.~\cite{huang2024survey} propose a taxonomy separating \emph{factuality} hallucination (conflict with world knowledge) from \emph{faithfulness} hallucination (conflict with the provided source), and survey detection and mitigation approaches, as do Zhang et al.~\cite{zhang2023sirens} and Alansari and Luqman~\cite{alansari2025survey}. Resume fabrication in our setting is a faithfulness failure with an unusually well-defined ground truth (the original resume plus an explicit absence list), which is what enables claim-level measurement rather than reference-free estimation.

\paragraph{Error propagation in multi-stage systems.}
Recent work shows that errors in multi-agent and multi-stage LLM systems propagate through message and context dependencies and can amplify into system-level failures~\cite{spark2026,jamshidi2026cascade}, and that pipeline architectures compound per-stage errors into end-to-end error rates exceeding any single component's~\cite{kotte2026pasc}. Our study contributes controlled measurements at a specific, deployable intervention point, a checkpoint between generation stages, rather than modeling propagation dynamics in the abstract.

\paragraph{Human oversight and its limits.}
HITL is a standard prescription in AI-governance frameworks and surveys~\cite{hitl2026review}, yet the empirical literature warns that oversight can become procedural rather than substantive~\cite{green2019principles}, that humans miscalibrate trust in fluent automated output~\cite{lee2004trust}, and that interface interventions such as cognitive forcing functions are needed to keep review engaged~\cite{bucinca2021trust}. Closest to our design, Zhu et al.~\cite{zhu2026attention} report a pre-specified factorial experiment in AI-assisted social-science research where an unconstrained multi-agent baseline produced critical failures in 72\% of runs and structured human-oversight architecture restored reliability. We complement this line with a fabrication-specific, claim-level measurement and with a characterization of the residual, namely what a protocol-following reviewer still misses, which prior HITL evaluations rarely quantify.

\paragraph{Guardrails.}
Mitigations range from external guardrail models (e.g., Llama Guard~\cite{inan2023llamaguard}, NeMo Guardrails~\cite{rebedea2023nemo}) to prompt-level safety instructions, which are known to be fragile: in-context guardrails can be overridden or simply not followed, and prompt engineering alone is generally regarded as insufficient as a safety control~\cite{dong2024position}. Our \Ctwo{} condition quantifies this insufficiency in a non-adversarial setting: even without any attacker, a strict factual-rules prompt left fabrication in half of outputs.

\paragraph{LLM-as-judge validity.}
LLM judges exhibit self-preference bias, favoring outputs from their own model or family~\cite{wataoka2024selfpref,pombal2026rubric,yang2026quantifying}. We therefore use a judge from a different vendor than the generation model, and validate it against blind human judgment on a stratified sample (\S\ref{sec:judgevalidity}).

\paragraph{Recruiting NLP.}
LLM pipelines for resume tailoring are already deployed as end-user tools; ResumeFlow~\cite{zinjad2024resumeflow} is a representative LLM pipeline that extracts JD and resume details and generates a role-specific resume, and motivates our content-preservation check (\S\ref{sec:metrics}). Our results speak directly to the safety properties this class of tool requires.

\section{Method}

\subsection{Benchmark construction}\label{sec:benchmark}

\paragraph{Resumes.}
Ten synthetic, PII-free resumes (R1--R10) with controlled variation in occupation, structure, information density, and specialty composition. Four resumes are multi-specialty with a controlled \emph{domain-distance gradient} (near: R1 $\to$ far: R7), enabling measurement of cross-domain contamination (e.g., graphic design experience rewritten as UX experience).

\paragraph{Ground truth.}
Each resume has a claim sheet: an enumeration of atomic claims (IDs C01\ldots) with type labels (METRIC, CRED, SKILL, SCOPE, ROLE, \ldots), plus an explicit absence list (IDs A01\ldots) of plausible-but-absent items (certifications, tools, domain experience, language levels) that serve as fabrication traps.

\paragraph{Job descriptions.}
Twenty JDs, two per resume: a \emph{proximate} JD ($\sim$75\% requirement overlap) and a \emph{stretch} JD ($\sim$45\%). Each JD embeds one trap requirement (rotating across certification / tool / domain / language-level types). Items intended for \emph{unprompted} fabrication measurement (e.g., Figma, Jira, JLPT) were excluded from all JDs.

\subsection{Conditions}

\begin{itemize}
  \item \textbf{\Cone{} (automated baseline)}: stages connected automatically, with a representative prompt containing neither fabrication-inducing language nor a guardrail. \Cone{} is deliberately non-adversarial; fabrication observed under it reflects typical developer prompting.
  \item \textbf{\Ctwo{} (prompt guardrail)}: identical pipeline; the stage-1 prompt is augmented with strict factual rules (no unsupported additions, no qualifier upgrades, no domain reinterpretation, no papering over gaps).
  \item \textbf{\Cthree{} (HITL)}: identical prompts to \Cone{} (isolating the human effect from the guardrail effect), with two human checkpoints: after resume improvement and after question generation. The pipeline halts; the reviewer corrects the output against the ground-truth claim sheet; the corrected text is consumed downstream. All edits are recorded as diffs with timestamps and content hashes.
\end{itemize}

A combined condition (C4 = guardrail + HITL) was deliberately deferred to future work to keep single-factor effects separable (\S\ref{sec:future}).

\subsection{Execution}

10 resumes $\times$ 2 JDs $\times$ 3 conditions $\times$ 3 repetitions = 180 runs, order randomized at the pair level (seed = 20260702). Generation model: \texttt{claude-sonnet-4-6}, temperature 0.7 (variance source across repetitions), no fallback models, fixed max-token budgets per stage. All stage inputs/outputs, prompt version IDs, and timestamps logged as JSONL; prompts under version control. Candidate answers were generated by a separate, condition-invariant LLM call referencing only the original resume.

\paragraph{Supplementary run (SUP).}
Stage 1 only, \texttt{claude-fable-5}, 20 pairs $\times$ 1 repetition. This model rejects the \texttt{temperature} parameter; default sampling was used.

\subsection{Reviewer protocol (\Cthree)}\label{sec:protocol}

A single reviewer (the first author) followed a fixed written protocol (Reviewer Protocol v1.1): a mandated checking order (numbers $\to$ credentials $\to$ qualifiers $\to$ tools/skills $\to$ domain boundaries $\to$ gaps $\to$ JD-requirement audit), an edit-scope rule (restore factuality only; no quality improvement; no additions), and pre-registered borderline decision rules. A reference AI (separate model family from the generation model) was permitted only for dictionary/search-style lookups of individual terms; delegating evaluation of pipeline output to any AI was prohibited, preserving the definition of \Cthree{} as a \emph{human} checkpoint. Review sessions were capped at approximately 2--2.5 hours to control for fatigue.

\paragraph{Protocol deviations.}
The reviewer's session memo records three checkpoint deviations, each identified at the time of occurrence: \texttt{R4\_B\_C3\_rep1 / stage1\_resume} and \texttt{R4\_B\_C3\_rep1 / stage2\_questions} were confirmed without edits being performed, and \texttt{R6\_A\_C3\_rep2 / stage2\_questions} was skipped. Per protocol, confirmed reviews were never retroactively edited; deviant runs are retained and disclosed (\S\ref{sec:deviations}). Only the stage-1 deviation (\texttt{R4\_B\_C3\_rep1}) affects the stage-1 fabrication analysis reported in \S\ref{sec:results}.

\subsection{Fabrication judgment}\label{sec:judgment}

All 260 stage-1 outputs were judged by an LLM judge from a different vendor (Gemini 3.5 Flash, temperature 0, prompt \texttt{judge-S1-v1}) to avoid self-preference bias: 240 from the main run (60 each for \Cone, \Ctwo, C3\_pre, and C3\_post, since \Cthree{} contributes both a pre-review and a post-review version of its 60 runs) plus 20 from SUP. The judge received the output, the original resume, and the ground-truth claim/absence sheets, and returned structured findings in five categories:

\begin{enumerate}
  \item \textbf{identity}: invented identity elements
  \item \textbf{absence}: hits on the explicit absence list (A-IDs)
  \item \textbf{qualifier\_drop}: dropped limiting qualifiers (e.g., ``basic'', ``prototype'', ``consecutive'')
  \item \textbf{altered}: modification of existing claims (scope/role/metric inflation)
  \item \textbf{new\_unsupported}: new claims with no grounding in the original resume
\end{enumerate}

Structured-output mode (\texttt{response\_mime\_type="application/json"}) caused indefinite non-response with this prompt and was not used; JSON was requested via instruction and parsed with code-fence stripping.

\subsection{Judge validity}\label{sec:judgevalidity}

Twenty items (stratified: \Cone:5, \Ctwo:5, C3\_pre:4, C3\_post:4, SUP:2; fixed seed) were manually judged blind by the reviewer against ground truth before unblinding the judge's verdicts. Pre-adjudication binary (fabrication present/absent) agreement: 17/20 (85\%). All three disagreements were embellishment-boundary cases in the \emph{new/altered} region; each was adjudicated by re-checking the flagged text against the original resume. Adjudication outcomes: one human miss (the judge was correct), one judge false positive, one judge false negative. Against the adjudicated reference, judge accuracy is 18/20 (90\%), with precision 15/16 (93.8\%) and recall 15/16 (93.8\%) at the item level. The boundary cases concentrate in the \emph{new/altered} region, the largest finding category (\S\ref{sec:results}), which we disclose as a validity consideration; the blind human reviewer also missed one boundary case (19/20), consistent with the reviewer-miss profile reported in \S\ref{sec:checkpoint}.

\paragraph{Non-independence disclosure.}
For C3\_post items, the manual validity reviewer is the same person who performed the \Cthree{} review being evaluated. Agreement figures for the C3\_post stratum should therefore be read with this caveat. Table~\ref{tab:strata} reports per-stratum agreement.

\begin{table}[t]
\centering
\caption{Judge validity by stratum: pre-adjudication agreement and post-adjudication judge accuracy.}
\label{tab:strata}
\begin{tabular}{lcccl}
\toprule
Stratum & Items & Agree (pre-adj.) & Judge correct (post-adj.) & Notes \\
\midrule
\Cone & 5 & 5/5 & 5/5 & \\
\Ctwo & 5 & 3/5 & 4/5 & one human miss, one judge false positive \\
C3\_pre & 4 & 3/4 & 3/4 & one judge false negative \\
C3\_post & 4 & 4/4 & 4/4 & reviewer = verifier (non-independent) \\
SUP & 2 & 2/2 & 2/2 & \\
\bottomrule
\end{tabular}
\end{table}

All three disagreements fell in independent strata (\Ctwo, C3\_pre), that is, in low-density strata where item-level verdicts hinge on single embellishment-boundary findings rather than on unambiguous fabrications. The non-independent C3\_post stratum showed perfect agreement, which we report without leaning on it.

\subsection{Metrics and analysis}\label{sec:metrics}

The primary metric is finding density (findings per output). Binary item-level fabrication rate (any finding) is reported as a secondary metric; it saturates near ceiling in \Cone/C3\_pre (96.7\%) and compresses between-condition differences. Content preservation is measured as claim retention: the judge (prompt \texttt{judge-S1-quality-v1}, same model and settings as \S\ref{sec:judgment}) marks each ground-truth claim ID as retained (substance present, rewording allowed) or dropped; retention = retained/total per output. Usefulness is a 1--5 rubric (JD alignment, specificity, readability) judged independently of factual accuracy, which is measured separately. Condition comparisons use pair-level rates ($n$ = 20 resume$\times$JD pairs, averaging over repetitions) with exact two-sided sign tests; 95\% CIs are Wilson intervals. The unit-of-analysis choice (pair-level) accounts for clustering of repetitions within pairs.

\section{Results}\label{sec:results}

\begin{figure}[t]
\centering
\includegraphics[width=\textwidth]{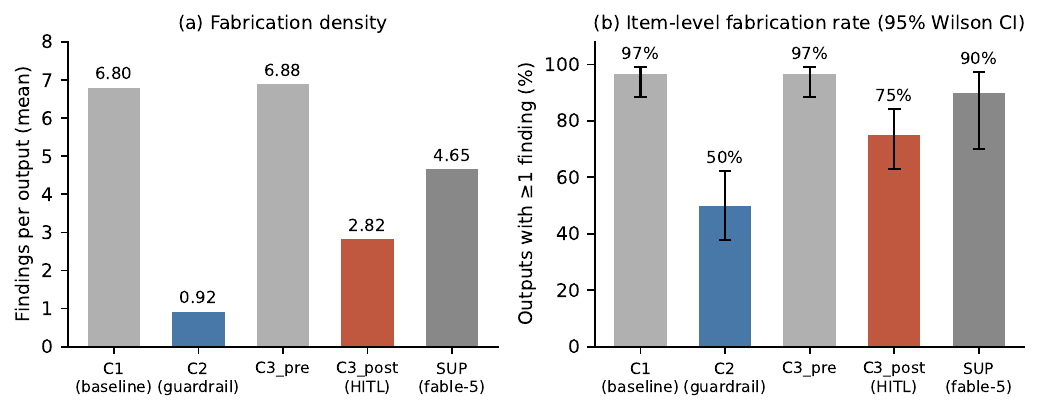}
\caption{Fabrication by condition. (a) Finding density (mean findings per output). (b) Item-level fabrication rate with 95\% Wilson intervals.}
\label{fig:conditions}
\end{figure}

\subsection{Fabrication is the default behavior (RQ1)}

\Cone{} produced at least one unsupported claim in 58/60 outputs (96.7\%, CI [88.6, 99.1]), with a mean of 6.80 findings per output (408 findings in total). The largest category was \emph{new\_unsupported} (237 findings, 58\%), followed by \emph{altered} (81) and \emph{absence-list hits} (57). Identity fabrication occurred in 6 outputs. C3\_pre, generated with the same prompt in independent runs, replicated this profile almost exactly (96.7\%, 6.88/output), passing the design sanity check (pair-level difference 0.0\%, $p = 1.0$).

\subsection{Guardrails help substantially but are insufficient (RQ2)}

\Ctwo{} reduced finding density by 86\% (6.80 to 0.92 per output) and eliminated two categories entirely (identity: 0, qualifier\_drop: 0). Item-level fabrication fell from 96.7\% to 50.0\% (pair-level sign test: 15/0, $p = .0001$). The residual was dominated by \emph{new\_unsupported} findings (40), despite the guardrail explicitly prohibiting additions; half of guardrail-protected outputs still contained at least one fabrication. Prompt-level mitigation alone therefore does not yield a deployable safety property.

\subsection{The human checkpoint: large effect, systematic residual (RQ3)}\label{sec:checkpoint}

\begin{figure}[t]
\centering
\includegraphics[width=0.8\textwidth]{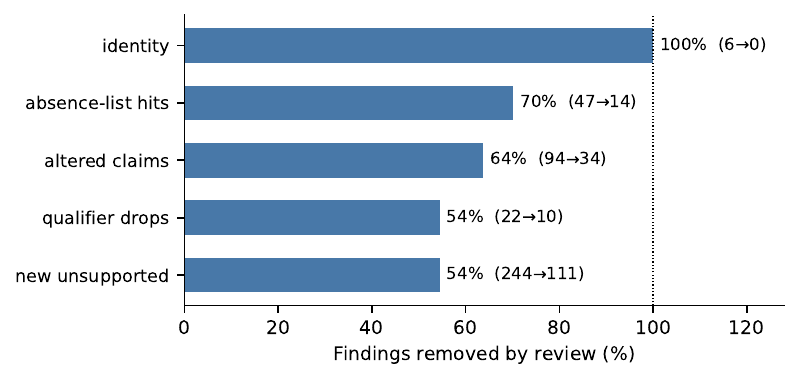}
\caption{Findings removed by the \Cthree{} review, by category (counts pre $\to$ post in parentheses).}
\label{fig:removal}
\end{figure}

The checkpoint reduced finding density by 59\% (6.88 to 2.82 per output) and item-level fabrication from 96.7\% to 75.0\% (pair-level sign test: 9/1, $p = .022$). Run-level dynamics ($n$ = 60 matched pre/post) are shown in Table~\ref{tab:prepost}; category-level removal rates (Table~\ref{tab:removal}, Figure~\ref{fig:removal}) reveal a clear gradient.

\begin{table}[t]
\centering
\caption{\Cthree{} run-level dynamics: fabrication status before and after review ($n$ = 60).}
\label{tab:prepost}
\begin{tabular}{lcc}
\toprule
 & Post: fab & Post: clean \\
\midrule
Pre: fab & 44 & 14 \\
Pre: clean & 1 & 1 \\
\bottomrule
\end{tabular}
\end{table}

\begin{table}[t]
\centering
\caption{Category-level finding counts before and after review, and removal rates.}
\label{tab:removal}
\begin{tabular}{lccc}
\toprule
Category & Pre & Post & Removal \\
\midrule
identity & 6 & 0 & 100\% \\
absence-list hits & 47 & 14 & 70.2\% \\
altered & 94 & 34 & 63.8\% \\
qualifier\_drop & 22 & 10 & 54.5\% \\
new\_unsupported & 244 & 111 & 54.5\% \\
\bottomrule
\end{tabular}
\end{table}

The reviewer in this study caught every flagrant fabrication (invented identities) and most trap hits, but subtle qualifier drops and plausible new claims survived review roughly half the time. One review introduced a new finding (pre-clean $\to$ post-fab, 1 run). We interpret this gradient as a specification of what reviewer-support tooling should provide: automated diff highlighting against source claims and qualifier-preservation checks, which target the classes of error an unaided reviewer under time pressure is most likely to miss (\S\ref{sec:limitations}).

\subsection{Guardrail vs.\ checkpoint}\label{sec:gvsc}

\Ctwo{} vs.\ C3\_post: item-level rates 50.0\% vs.\ 75.0\% (sign test 5/11, $p = .21$, n.s.); densities 0.92 vs.\ 2.82. On fabrication metrics alone the guardrail condition is at least as clean, and \S\ref{sec:quality} shows that this is not an artifact of content loss. The empirical case for the checkpoint therefore rests not on aggregate counts but on (i) the categorical guarantee no prompt achieved (identity removal 100\%), (ii) complementary failure modes (\Ctwo's residual is plausible additions the model generates \emph{despite instruction}; \Cthree's residual is what a reviewer misses), and (iii) slightly better claim retention and JD alignment than the guardrail (\S\ref{sec:quality}). The combined condition (C4) is the direct test of this complementarity.

\subsection{Newer model, same problem (RQ4)}

SUP (\texttt{claude-fable-5}, baseline prompt) showed 90.0\% item-level fabrication (CI [69.9, 97.2]) and 4.65 findings per output, with \emph{new\_unsupported} again the dominant category (54/93). Identity fabrication: 0. A generation of model progress changes the profile (no invented identities) but not the conclusion: unguarded pipelines fabricate by default.

\subsection{Content preservation and usefulness}\label{sec:quality}

Claim retention was near ceiling in all conditions (\Cone{} 97.6\%, \Ctwo{} 99.2\%, C3\_post 99.4\%, SUP 99.8\%). The ``safe but empty'' hypothesis for the guardrail is not supported: \Ctwo's low finding counts are not explained by content loss. The only significant retention contrast is \Cone{} vs C3\_post (97.6\% vs 99.4\%, $p = .039$): the automated baseline drops slightly more source claims than the reviewed pipeline, consistent with fabrication displacing grounded content.

Usefulness scores were high everywhere (4.5--5.0). \Cone{} scored nominally highest (mean 4.91), with a significant jd\_alignment advantage over \Ctwo{} (4.88 vs 4.55, $p = .022$). This advantage must be read jointly with the fabrication results: the baseline achieves alignment in part by inventing JD-matching qualifications (237 unsupported new claims), so its usefulness edge is partly fabricated alignment and is unusable in any honest deployment. Between the two mitigations, usefulness does not differ significantly (\Ctwo{} 4.77 vs C3\_post 4.86, $p = .34$); the checkpoint recovers some of the alignment the guardrail forgoes (jd\_alignment 4.75 vs 4.55) without the baseline's fabrication.

Both mitigations therefore preserve the deliverable: safety gains were not purchased with content loss or material usefulness loss.

\subsection{Handling of protocol deviations}\label{sec:deviations}

One deviant run (\texttt{R4\_B\_C3\_rep1}, unreviewed at the stage-1 checkpoint; \S\ref{sec:protocol}) is included in the primary analysis. Its direction of bias is conservative: the unreviewed run contributes its uncorrected fabrications to C3\_post, so the reported checkpoint effect is, if anything, slightly understated. At 1 of 60 \Cthree{} runs, the magnitude is negligible; we disclose rather than exclude. The two stage-2 deviations affect only the question-stage analysis.

\subsection{Induced vs.\ unprompted fabrication (trap analysis)}\label{sec:traps}

Each JD embeds one trap requirement, a plausible skill, credential, domain, or language requirement absent from the resume (\S\ref{sec:benchmark}). For 19 of the 20 pairs the trap is an explicit absence-list item (A-ID) and thus directly measurable in the judgment schema; the remaining pair's trap (a language-level requirement) lies outside the absence list by design and is excluded with this disclosure. Trap capture (the output implies the candidate meets the baited requirement) separates \emph{induced} fabrication from \emph{unprompted} fabrication (absence-list hits with no corresponding JD requirement); see Table~\ref{tab:traps}.

\begin{table}[t]
\centering
\caption{Trap capture and unprompted absence-list hits by condition (19 measurable pairs).}
\label{tab:traps}
\begin{tabular}{lccccc}
\toprule
 & \Cone & \Ctwo & C3\_pre & C3\_post & SUP \\
\midrule
Trap capture rate & 47.4\% & 5.3\% & 47.4\% & 1.8\% & 10.5\% \\
Unprompted hits / output & 0.46 & 0.00 & 0.35 & 0.21 & 0.21 \\
\bottomrule
\end{tabular}
\end{table}

Two asymmetries stand out. First, the checkpoint was the strongest defense against \emph{baited} fabrication (1.8\%, 1/57, versus 5.3\% for the guardrail), plausibly because the reviewer protocol's final step is an explicit JD-requirement audit; this constitutes the checkpoint's second near-categorical guarantee alongside identity removal (\S\ref{sec:checkpoint}). Second, the guardrail, but not the reviewer, eliminated \emph{unprompted} absence-list hits entirely (0.00 vs.\ 0.21 per output), while remaining more permeable to JD-baited items. The two mechanisms are strongest on opposite halves of the same phenomenon, sharpening the complementarity case (\S\ref{sec:gvsc}). Among measured traps, the highest-yield bait was the plausible domain upgrade predicted at design time (e.g., marketing experience recast as employer-branding strategy), confirming that traps adjacent to real experience are the most readily taken.

\subsection{Cross-domain contamination along the specialty-distance gradient (exploratory)}\label{sec:contamination}

Four resumes place two specialties at increasing domain distance: R1 (design + frontend, near), R9 (recruiting + marketing, near-mid), R3 (graphic design + IT program management, mid-far), R7 (German legal practice + localization, far). Pooling \Cone{} and C3\_pre (identical prompt; 12 outputs per resume), contamination-type alterations (ROLE + SCOPE) increase monotonically with domain distance: 12 (R1), 13 (R9), 23 (R3), and 36 (R7). Multi-specialty resumes also carried higher overall finding density than the six single-specialty resumes (7.92 vs.\ 6.12 findings/output, +29\%). With one resume per distance level this analysis is exploratory and no tests are performed, but the pattern is consistent and the mechanism is visible in the outputs: R1's designer acquires an invented hybrid title (``Design Engineer''), R3's design-career achievements are re-scoped as technical program governance, and R7, the farthest pair, shows the heaviest fusion, with legal-practice experience rewritten into localization program leadership. In practical terms, candidates whose careers span distant domains face materially higher fabrication exposure from automated rewriting; career changers and multi-specialty professionals, the population such tools most attract, are the most affected. A confirmatory study with multiple resumes per distance level is future work (\S\ref{sec:future}).

\section{Discussion}

\paragraph{Both mitigations are effective; neither alone reaches deployable safety.}
The central practical question for hiring pipelines is what it takes to make automated rewriting safe. Prompting alone does not suffice, as the best guardrail still shipped fabrication in half of outputs; a human checkpoint alone does not suffice either, as 75\% of reviewed outputs retained at least one finding. The two interventions are, however, strong in different regions of the failure distribution. The guardrail cut finding density by 86\% and eliminated unprompted absence-list hits and qualifier inflation entirely; the checkpoint provided guarantees no prompt achieved in our data, with zero identity fabrications and 1.8\% trap capture after review. In deployment contexts where a single fabricated credential is a serious harm, and hiring is such a context because the fabrication becomes the premise of a real interview, the empirically supported architecture is layered: guardrails to shrink the fabrication volume at negligible cost, and a human checkpoint to provide the categorical guarantees.

\paragraph{Residual reviewer error as a tooling specification.}
The removal gradient (100\% for identities, roughly 55\% for qualifier drops and new claims) maps directly onto interface design: the reviewer had a claim sheet and a fixed checking order, but no automated support for aligning output sentences to source claims. The two weakest categories, dropped qualifiers and plausible new claims, are the categories where machine assistance such as diff-to-source alignment and qualifier tracking is inexpensive while unaided human attention is costly. Checkpoint effectiveness depends on checkpoint design; this study measured an unaided, text-only, single-reviewer configuration, which is plausibly a lower bound.

\paragraph{Combining the two mechanisms.}
Guardrails and checkpoints fail on different parts of the fabrication distribution. A layered architecture follows naturally: guardrails shrink the volume the reviewer must inspect (the reduction from 6.80 to 0.92 findings per output corresponds to an 86\% reduction in review burden), and the reviewer provides the categorical guarantees that prompts did not achieve. The combined condition (C4) is the direct next experiment (60 runs, same benchmark).

\paragraph{Reviewer fallibility as data.}
We report protocol deviations and one review-introduced finding rather than excluding them. In a study of human oversight, the behavior of the study's own reviewer, including errors, is part of the measured phenomenon, and the residual-error profile in \S\ref{sec:checkpoint} is informative for checkpoint design for that reason.

\section{Limitations}\label{sec:limitations}

\begin{enumerate}
  \item \textbf{Single reviewer ($n$ = 1).} All removal rates and error profiles describe one reviewer under one protocol. Between-reviewer variance, training effects, and fatigue dynamics are unmeasured; generalization requires a multi-reviewer study. Claims in this paper are therefore about the presence and size of a checkpoint effect, not about human reviewers in general.
  \item \textbf{Judge validity is estimated on 20 items.} Adjudicated accuracy 90\% (precision/recall 93.8\%) is strong but carries a wide interval at $n$ = 20; residual judge error concentrates in the embellishment boundary of the largest finding category (new/altered), so category-level counts there carry the most measurement uncertainty.
  \item \textbf{C3\_post validity non-independence} (reviewer = validity verifier for that stratum), disclosed in \S\ref{sec:judgevalidity}.
  \item \textbf{Single generation model / single domain} (hiring, English-language resumes). The SUP run probes model generality only at baseline.
  \item \textbf{Synthetic benchmark.} PII-free construction was an ethical requirement but may understate the messiness of real resumes.
  \item \textbf{Quality metrics show ceiling effects.} Usefulness scores cluster at 4.5--5.0 and retention at 97--100\%, indicating judge leniency; the quality judge was not separately validated against human ratings. These metrics support the narrow claim made from them (no material content or usefulness loss under either mitigation) but would not resolve fine-grained quality differences.
  \item \textbf{Two-checkpoint design, stage-1 results only.} \Cthree{} includes a second checkpoint after question generation, but all results reported here are measured on stage-1 outputs; stage-2 checkpoint effects are analyzed separately.
\end{enumerate}

\section{Future Work}\label{sec:future}

C4 (guardrail + HITL, 60 runs); multi-reviewer replication with tooling-assisted vs.\ unaided review as a factor; propagation analysis through stages 2--3 (question generation and feedback); a confirmatory cross-domain contamination study with multiple resumes per distance level, extending the exploratory gradient result (\S\ref{sec:contamination}).

\section{Conclusion}

In a controlled 180-run comparison, fully automated resume-improvement pipelines fabricated in 96.7\% of outputs. Both mitigations proved effective in complementary ways. Prompt guardrails cut fabrication density by 86\% and eliminated unprompted absence-list hits entirely; a single human checkpoint eliminated all identity-level fabrication, reduced JD-baited trap capture to 1.8\%, and cut density by 59\%. Neither intervention sacrificed content or usefulness. For multi-stage LLM pipelines in hiring, the evidence supports deploying both, with guardrails as a low-cost first layer and human checkpoints as the source of categorical guarantees; the measured residual errors of each indicate the reviewer-support tooling and the combined-condition experiment that should follow.

\section*{Acknowledgments}

We thank Ludovica Erhardt for her generous assistance with the construction of the synthetic data benchmark, and for her sustained feedback and support throughout this project.

\section*{Reproducibility}

Harness, prompts (versioned), benchmark (resumes, claim sheets, absence lists, JDs), judge prompt, seeds, and full JSONL logs are available upon request.

\section*{Ethics statement}

All resumes and job descriptions are fully synthetic and PII-free; no real candidates' data were used, and no fabricated identity details generated during the experiment are reproduced in this paper.

\bibliographystyle{unsrtnat}
\bibliography{references}

@article{huang2024survey,
  author  = {Huang, Lei and Yu, Weijiang and Ma, Weitao and Zhong, Weihong
             and Feng, Zhangyin and Wang, Haotian and Chen, Qianglong
             and Peng, Weihua and Feng, Xiaocheng and Qin, Bing and Liu, Ting},
  title   = {A Survey on Hallucination in Large Language Models: Principles,
             Taxonomy, Challenges, and Open Questions},
  journal = {{ACM} Transactions on Information Systems},
  volume  = {43},
  number  = {2},
  pages   = {42:1--42:55},
  year    = {2025},
  doi     = {10.1145/3703155},
  note    = {arXiv:2311.05232}
}

@article{zhang2023sirens,
  author  = {Zhang, Yue and Li, Yafu and Cui, Leyang and Cai, Deng
             and Liu, Lemao and Fu, Tingchen and Huang, Xinting
             and Zhao, Enbo and Zhang, Yu and Xu, Chen and Chen, Yulong
             and Wang, Longyue and Luu, Anh Tuan and Bi, Wei and Shi, Freda
             and Shi, Shuming},
  title   = {Siren's Song in the {AI} Ocean: A Survey on Hallucination in
             Large Language Models},
  journal = {arXiv preprint arXiv:2309.01219},
  year    = {2023}
}

@article{alansari2025survey,
  author  = {Alansari, Aisha and Luqman, Hamzah},
  title   = {Large Language Models Hallucination: A Comprehensive Survey},
  journal = {arXiv preprint arXiv:2510.06265},
  year    = {2025}
}

@article{spark2026,
  author  = {Xie, Yizhe and Zhu, Congcong and Zhang, Xinyue and Zhu, Tianqing
             and Ye, Dayong and Qi, Minfeng and Chen, Huajie and Zhou, Wanlei},
  title   = {From Spark to Fire: Modeling and Mitigating Error Cascades in
             {LLM}-Based Multi-Agent Collaboration},
  journal = {arXiv preprint arXiv:2603.04474},
  year    = {2026}
}

@article{jamshidi2026cascade,
  author  = {Jamshidi, Saeid and Moradi Dakhel, Arghavan
             and Nafi, Kawser Wazed and Khomh, Foutse},
  title   = {Hallucination Cascade: Analyzing Error Propagation in
             Multi-Agent {LLM} Systems},
  journal = {arXiv preprint arXiv:2606.07937},
  year    = {2026}
}

@article{kotte2026pasc,
  author  = {Kotte, Varun},
  title   = {{PASC}: Pipeline-Aware Conformal Prediction with Joint Coverage
             Guarantees for Multi-Stage {NLP} and {LLM} Pipelines},
  journal = {arXiv preprint arXiv:2605.18812},
  year    = {2026}
}

@article{hitl2026review,
  author  = {Lazaros, Konstantinos and Vrahatis, Aristidis G.
             and Kotsiantis, Sotiris},
  title   = {Human-in-the-Loop Artificial Intelligence: A Systematic Review of
             Concepts, Methods, and Applications},
  journal = {Entropy},
  volume  = {28},
  number  = {4},
  pages   = {377},
  year    = {2026},
  doi     = {10.3390/e28040377}
}

@article{green2019principles,
  author  = {Green, Ben and Chen, Yiling},
  title   = {The Principles and Limits of Algorithm-in-the-Loop Decision
             Making},
  journal = {Proceedings of the {ACM} on Human-Computer Interaction},
  volume  = {3},
  number  = {CSCW},
  pages   = {50:1--50:24},
  year    = {2019},
  doi     = {10.1145/3359152}
}

@article{lee2004trust,
  author  = {Lee, John D. and See, Katrina A.},
  title   = {Trust in Automation: Designing for Appropriate Reliance},
  journal = {Human Factors},
  volume  = {46},
  number  = {1},
  pages   = {50--80},
  year    = {2004},
  doi     = {10.1518/hfes.46.1.50_30392}
}

@article{bucinca2021trust,
  author  = {Bu{\c{c}}inca, Zana and Malaya, Maja Barbara
             and Gajos, Krzysztof Z.},
  title   = {To Trust or to Think: Cognitive Forcing Functions Can Reduce
             Overreliance on {AI} in {AI}-Assisted Decision-Making},
  journal = {Proceedings of the {ACM} on Human-Computer Interaction},
  volume  = {5},
  number  = {CSCW1},
  pages   = {188:1--188:21},
  year    = {2021},
  doi     = {10.1145/3449287}
}

@article{zhu2026attention,
  author  = {Zhu, Chen and Wang, Xiaolu and Zhang, Weilong},
  title   = {({H}uman) Attention Is (Still) All You Need: Human Oversight Makes
             {AI}-Assisted Social Science Reliable},
  journal = {arXiv preprint arXiv:2606.12848},
  year    = {2026}
}

@article{inan2023llamaguard,
  author  = {Inan, Hakan and Upasani, Kartikeya and Chi, Jianfeng
             and Rungta, Rashi and Iyer, Krithika and Mao, Yuning
             and Tontchev, Michael and Hu, Qing and Fuller, Brian
             and Testuggine, Davide and Khabsa, Madian},
  title   = {{Llama Guard}: {LLM}-Based Input-Output Safeguard for
             Human-{AI} Conversations},
  journal = {arXiv preprint arXiv:2312.06674},
  year    = {2023}
}

@inproceedings{rebedea2023nemo,
  author    = {Rebedea, Traian and Dinu, Razvan
               and Sreedhar, Makesh Narsimhan and Parisien, Christopher
               and Cohen, Jonathan},
  title     = {{NeMo Guardrails}: A Toolkit for Controllable and Safe {LLM}
               Applications with Programmable Rails},
  booktitle = {Proceedings of the 2023 Conference on Empirical Methods in
               Natural Language Processing: System Demonstrations},
  pages     = {431--445},
  year      = {2023},
  doi       = {10.18653/v1/2023.emnlp-demo.40},
  note      = {arXiv:2310.10501}
}

@inproceedings{dong2024position,
  author    = {Dong, Yi and Mu, Ronghui and Jin, Gaojie and Qi, Yi
               and Hu, Jinwei and Zhao, Xingyu and Meng, Jie
               and Ruan, Wenjie and Huang, Xiaowei},
  title     = {Position: Building Guardrails for Large Language Models
               Requires Systematic Design},
  booktitle = {Proceedings of the 41st International Conference on Machine
               Learning},
  series    = {Proceedings of Machine Learning Research},
  volume    = {235},
  pages     = {11375--11394},
  year      = {2024}
}

@article{wataoka2024selfpref,
  author  = {Wataoka, Koki and Takahashi, Tsubasa and Ri, Ryokan},
  title   = {Self-Preference Bias in {LLM}-as-a-Judge},
  journal = {arXiv preprint arXiv:2410.21819},
  year    = {2024}
}

@article{pombal2026rubric,
  author  = {Pombal, Jos{\'e} and Rei, Ricardo and Martins, Andr{\'e} F. T.},
  title   = {Self-Preference Bias in Rubric-Based Evaluation of Large
             Language Models},
  journal = {arXiv preprint arXiv:2604.06996},
  year    = {2026}
}

@article{yang2026quantifying,
  author  = {Yang, Jinming and Hu, Zheng and Qiu, Chuxian and Deng, Zhenyu
             and Jiao, Xinshan and Zhou, Tao},
  title   = {Quantifying and Mitigating Self-Preference Bias of {LLM} Judges},
  journal = {arXiv preprint arXiv:2604.22891},
  year    = {2026}
}

@inproceedings{zinjad2024resumeflow,
  author    = {Zinjad, Saurabh Bhausaheb and Bhattacharjee, Amrita
               and Bhilegaonkar, Amey and Liu, Huan},
  title     = {{ResumeFlow}: An {LLM}-Facilitated Pipeline for Personalized
               Resume Generation and Refinement},
  booktitle = {Proceedings of the 47th International {ACM} {SIGIR} Conference
               on Research and Development in Information Retrieval},
  pages     = {2781--2785},
  year      = {2024},
  doi       = {10.1145/3626772.3657680},
  note      = {arXiv:2402.06221}
}

\end{document}